%% file: lubini.tex
\documentclass[useAMS,usenatbib]{mn2e}

\usepackage{amsmath,amssymb,amsfonts}
\usepackage{graphicx}
\usepackage{aas_macros}
\usepackage{color}

\newcommand{\abs}[1]{\left|#1\right|}
\newcommand{\vc}[1]{\boldsymbol{#1}}
\newcommand{\mat}[1]{{\bf #1}}
\newcommand{\dd}{\mathrm{d}}
\newcommand{\R}{\mathbb{R}}
\newcommand{\kinf}{\tilde{\kappa}}
\newcommand{\kref}{{\kappa_\mathrm{ref}}}
\newcommand{\z}[1]{{z_\mathrm{#1}}}
\newcommand{\Om}[1]{\Omega_\mathrm{#1}}
\newcommand{\da}[1]{D_{\mathrm{#1}}}
\newcommand{\dr}[1]{\Delta_{\mathrm{#1}}}
\newcommand{\dof}{{n_\mathrm{dof}}}

\newcommand{\secref}[1]{Section~\ref{#1}}
\newcommand{\eqnref}[1]{Eq.~\eqref{#1}}
\newcommand{\eqnsref}[1]{Eqs.~\eqref{#1}}
\newcommand{\figref}[1]{Figure~\ref{#1}}

\begin{document}

\title[Cosmography in Free-Form Gravitational Lensing]
	{Cosmological Parameter Determination in Free-Form Strong Gravitational Lens Modeling}

\author[M.~Lubini et al.]
	{M.~Lubini$^{1}$\thanks{E-mail: lubini@physik.uzh.ch},
	M.~Sereno$^{2,3}$, J.~Coles$^{1}$, Ph.~Jetzer$^{1}$ and P.~Saha$^{1}$\\
	$^{1}$Institut f\"{u}r Theoretische Physik, Universit\"{a}t Z\"{u}rich,
	Winterthurerstrasse 190, 8057 Z\"{u}rich, Switzerland.\\
	$^{2}$Dipartimento di Scienza Applicata e Tecnologia, Politecnico di Torino, Corso Duca degli
	Abruzzi 24, 10129 Torino, Italia\\
	$^{3}$INFN, Sezione di Torino, Via Pietro Giuria 1, 10125 Torino, Italia}

\maketitle

\label{firstpage}

\begin{abstract}
We develop a novel statistical strong lensing approach to probe the cosmological
parameters by exploiting multiple redshift image systems behind galaxies or
galaxy clusters. The method relies on free-form mass inversion of strong lenses
and does not need any additional information other than gravitational lensing.
Since in free-form lensing the solution space is a high-dimensional convex
polytope, we consider Bayesian model comparison analysis to infer the
cosmological parameters. The volume of the solution space is taken as a tracer
of the probability of the underlying cosmological assumption. In contrast to
parametric mass inversions, our method accounts for the mass-sheet degeneracy,
which implies a degeneracy between the steepness of the profile and the
cosmological parameters. Parametric models typically break this degeneracy,
introducing hidden priors to the analysis that contaminate the inference of the
parameters. We test our method with synthetic lenses, showing that it is able to
infer the assumed cosmological parameters. Applied to the CLASH
clusters, the method might be competitive with other probes.
\end{abstract}


\section{Introduction}
\label{sec:introduction}

Estimates of the matter/energy content of the Universe have reached
uncertainties of only a few percent through the combined analysis of the
anisotropy measurements of the CMB \citep{komatsu11-WMAP7,ade13-Planck}, the
observation of the baryon acoustic oscillations (BAO) in the distribution of
galaxies \citep{percival10}, and the luminosity distance of Type Ia supernovae
\citep{amanullah10,riess11}. Nonetheless, the precision cosmology era crucially
requires further independent methods in order to control systematic effects that
can plague some techniques and to break statistical degeneracies.

A unique tool is provided by gravitational lensing, which can furnish a rich
source of information about the underlying cosmological model. Gravitational
lensing relies on the angular diameter distances, which in turn depend on the
matter/energy content of the Universe. Particularly in galaxy clusters, the
identification of multiple gravitationally lensed background sources located at
different redshifts \citep[e.g.,][]{limousin07,richard09} supplies information
on the cosmological parameters. Galaxies as lenses can probe the cosmology too,
but only a few multiple source redshift lenses are presently known
\citep{bolton08}.

Unlike other probes, such as supernovae measurements, the cosmological
information contained in strong gravitational lenses is purely geometrical and
does not require any kind of calibration. Moreover, it probes cosmology in an
almost unexplored redshift range of around $z\sim 3-4$. Various other work
\citep[e.g.,][]{golse02,sereno02,sereno04,soucail04,gilmore09,daloisio11,zieser12}
has shown and investigated the ability of strong gravitational lensing to
determine the cosmological parameters in clusters of galaxies using parametric
lensing models. \citet{jullo10} constrained the mass distribution of the main
components of the galaxy cluster Abell 1689 and the dark energy equation of
state.

Parametric models assume a functional form for the lens mass distribution and
can be very efficient if all the cluster components are considered through
adequate mass profiles. These models, however, introduce hidden priors to the
analysis, as the assumed shape may unintentionally break possible degeneracies
between the cosmological parameters and the mass profile. For instance, the NFW
\citep{navarro96-NFW1,navarro97-NFW2} or isothermal density mass profiles can
both provide good fits to observed systems, but choosing one of the two
competitive profiles artificially breaks the mass-sheet degeneracy and biases
the analysis of cosmological parameters.

The mass-sheet degeneracy is one of the main limitations and source of
uncertainty in gravitational lensing mass estimation \citep{falco85,saha00}. A
significant endeavor to break this degeneracy in parametric models has been made
when modelling the mass profile of galaxies and galaxy clusters
\citep[e.g.,][]{suyu12,collett13,greene13,umetsu13}. Proper analyses of the
mass-sheet degeneracy should then be considered when investigating the
cosmological parameters.

Parametric models also demand deep knowledge of all the cluster components,
which can only be achieved though observations other than gravitational lensing,
e.g.,~optical for the position of the galactic halos, and X-ray for the
temperature and location of the intracluster medium
\citep{voit05,my:sereno10b,limousin12,sereno13}. Only lensing
clusters with deep multi-wavelength data sets can then be used to constrain
cosmological parameters through parametric models.

In this paper we apply a free-form approach to model the lens mass distribution.
This approach only requires the knowledge of the lensed image positions and
redshifts, and is more flexible than analytic models. Several different forms of
the basic strategy have been developed for clusters with given cosmological
parameters \citep{abdelsalam98a, abdelsalam98b, bradac04, bradac05, diego05,
	read07, liesenborgs07, deb08, 2008ApJ...681..814C}. In a given cosmology,
the presence of sources at different redshifts helps break lensing degeneracies.
In the present work, however, we do not fix the cosmology. Instead, we exploit
the multiple source redshifts to follow a formulation of Occam's razor for the
purpose of comparing competitive cosmological models. We consider a Bayesian
approach exploiting the statistical dispersion of the parameter space describing
the mass distribution to obtain information about the assumed cosmological
model.

In \secref{sec:framework} the cosmological information contained in strong
gravitational lensing, as well as the relevant degeneracies, are stated, whereas
the free-form lensing approach is laid out in \secref{sec:free-form}.
\secref{sec:method} presents the statistical method, which is based on Occam's
razor in Bayesian model comparison, whereas in \secref{sec:tests} we test the
method through synthetic lenses and show that we are able to account for the
mass-sheet degeneracy. The performance of the method in a realistic situation is
shown in \secref{sec:parameter-determination}. Conclusions are presented in
\secref{sec:conclusions}.

\section{Framework}
\label{sec:framework}

The basic relation in gravitational lensing is the lens equation
\citep{schneider92-SEF,schneider06}
\begin{equation}
\vc\beta = \vc\theta - \vc\alpha(\vc\theta) ,
\label{eq:lens-eq}
\end{equation}
which maps the observed image angular position $\vc\theta$ to the angular
position $\vc\beta$ of the source through the scaled deflection angle
\begin{equation}
\vc\alpha(\vc\theta)=\frac1\pi\int_{\R^2}\kappa(\vc\theta')\frac{\vc\theta-\vc\theta'}{\abs{\vc\theta-\vc\theta'}^2}\,\dd^2\theta' .
\label{eq:def-angle}
\end{equation}
The dimensionless surface mass density or convergence $\kappa$ is defined by
\begin{equation}
\kappa(\vc\theta)=\frac{\Sigma(\da{ol}\vc\theta)}{\Sigma_{\mathrm{crit}}}\quad\text{with}\quad\Sigma_{\mathrm{crit}}=\frac{c^2}{4\pi G}\frac{\da{os}}{\da{ls}\da{ol}} ,
\label{eq:kappa}
\end{equation}
where $\Sigma$ is the surface mass density of the lens and $\da{ol}$, $\da{os}$,
and $\da{ls}$ are the angular diameter distances between observer and lens,
observer and source, and lens and source, respectively.

We consider a model of universe with cold dark matter (CDM) whose accelerated
expansion is propelled by some form of dark energy. Assuming a dark energy with
a constant equation of state $w$ the angular diameter distance between the
redshifts $\z{a}$ and $\z{b}$ is \citep{weinberg72}
\begin{equation}
	\da{ab} = \frac{c}{(1+\z{b}) H_0 \sqrt{|\Omega_k|}}\,
	\mathcal{S}_k\left(\sqrt{|\Omega_k|}\int_\z{a}^\z{b}\frac{H_0}{H(z)}\dd z\right) ,
\end{equation}
where $\mathcal{S}_k(x)=x$, $\sin(x)$, or $\sinh(x)$ for a flat, closed, or open
universe, respectively. The Hubble parameter $H(z)$ is given by
\begin{equation}
	\frac{H(z)}{H_0} = \sqrt{\Om{m} (1+z)^3 + \Omega_k (1+z)^2 + \Om{de}(1+z)^{3(1+w)}} ,
\end{equation}
where $H_0=100\,h\,\mathrm{km\,s^{-1}\,Mpc^{-1}}$ is the Hubble constant with the
dimensionless Hubble parameter $h$ and the $\Omega$'s are the matter, curvature
$k$, and dark energy density parameters. The standard $\Lambda$CDM model with
cosmological constant $\Lambda$ is given by the special case with $w=-1$ and
$\Om{de}=\Omega_\Lambda$.

\subsection{Strong lensing cosmography}

The aim of this paper is to infer the cosmological parameters, which hereafter
are simply denoted by $\Omega$, through the distances in \eqnref{eq:kappa},
exploiting strong lensing observations in clusters of galaxies. Since what we
observe are the angular positions $\vc\theta$ of the lensed images, by means of
\eqnref{eq:lens-eq}, gravitational lensing only deduces $\kappa$ rather than the
true mass profile $\Sigma$ or any of the distances. However, when considering a
lensing object where multiple sources are observed, $\kappa$ can simultaneously
be inferred at different source redshifts. As we are dealing with multiple
source planes whereas the observer and lens redshifts are fixed, $\kappa$
depends only on the source redshift $\z{s}$. The source dependent part
\begin{equation}
	\dr{s} = \frac{\da{ls}}{\da{os}}
\end{equation}
can then be extracted from the convergence, which we rewrite as
\begin{equation}
	\kappa(\z{s}) = \dr{s}\, \kinf = \dr{s} \, \frac{\kref}{\dr{ref}} .
\label{eq:kinf}
\end{equation}
$\kinf$ can be interpreted as the convergence for a source geometrically at
infinity, i.e.~where $\dr{s}=1$, or alternatively one could consider the
convergence $\kref$ at some fixed reference source redshift $\z{ref}$.

\begin{figure*}
\centering
\includegraphics[width=\textwidth]{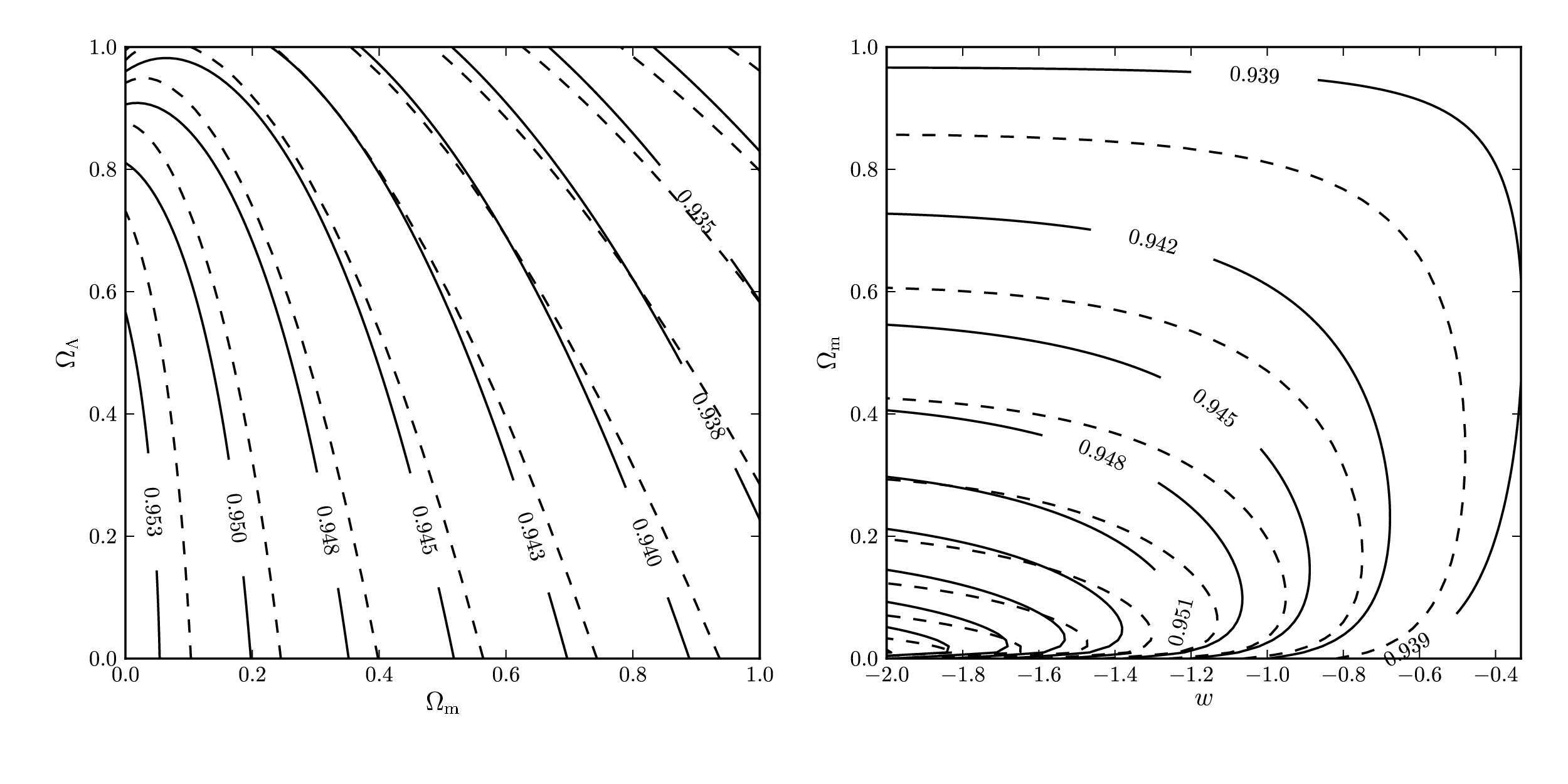}

\caption{Isodensity contours for $\Xi(\z{1},\z{2},\Omega)$ with source redshifts
	$(\z{1},\z{2})=(1.5,2.5)$ (solid lines) and $(\z{1},\z{2})=(2.0,3.0)$
	(dashed lines) are shown in the plane $\Om{m}-\Om{\Lambda}$ with $w=-1$
	(\emph{left panel}) and $w-\Om{m}$ with $\Omega_k = 0$ (\emph{right panel}).
	The redshift of the lens is $\z{l}=0.2$. In the $\Om{m}-\Om{\Lambda}$ plane
	the contours nearly go in the direction of constant curvature
	$\Omega_k=1-\Om{m}-\Om{\Lambda}=\mathrm{constant}$. This implies that
	gravitational lensing is particularly sensitive to the curvature of the
	universe. Along the isodensity contours the cosmological information is
	completely degenerate, but one can break this degeneracy by adding probes
	with different source redshifts.}

\label{fig:om-ol-w-contours}
\end{figure*}

In the case of a single source plane, $\kinf$ is completely degenerate with the
distance ratio $\dr{s}$, since gravitational lensing is only able to infer
$\kappa$ and both $\kinf$ and $\dr{s}$ in \eqnref{eq:kinf} are unknown.
Consequently, we can not constrain the cosmological parameters $\Omega$
contained in $\dr{s}$ by exploiting only a single image system. Additional
information on the cluster mass distribution is needed to break the degeneracy
between $\kinf$ and $\dr{s}$. This information is, for example, given by
dynamical analyses from optical observations of the velocity dispersion of the
cluster galaxies \citep[e.g.,][]{wojtak07}, or from X-ray observations which
reveal the luminosity, temperature, and location of the intracluster medium
\citep[e.g.,][]{vikhlinin06}.

The information from additional image systems can break this degeneracy, too. By
means of a second source plane at redshift $z_2$ gravitational lensing infers
$\kappa(z_2)$. Combining inferences from two source planes at the
redshifts $z_1$ and $z_2$ one obtains
\begin{equation}
	\frac{\kappa(z_1)}{\kappa(z_2)} =
	\frac{\dr{1}(\Omega)}{\dr{2}(\Omega)} =:
	\Xi(z_1,z_2,\Omega) ,
\label{eq:ratios-ratio}
\end{equation}
where the dependency on the cosmological parameters $\Omega$ is explicitly
given. This ratio does not depend on the lens mass distribution $\kinf$. By
comparing image positions of lensed sources at different redshifts we can then
construct a cosmological probe based on the ratio of distance ratios
$\Xi(z_1,z_2,\Omega)$. \figref{fig:om-ol-w-contours} shows the isodensity
contours of $\Xi(\z{1},\z{2},\Omega)$ for different redshift values in the
$\Om{m}-\Om{\Lambda}$ plane, as well as in the $w-\Om{m}$ plane, where a flat
universe is assumed. The cosmological information is completely degenerate
along the contours, where the value for $\Xi$ is constant. We can break this
degeneracy by combining data from three or more source planes.

In the left panel the degeneracy nearly goes in the direction of constant
curvature, meaning that gravitational lensing is particularly sensitive to the
curvature of the universe.

\subsection{Hidden priors}
\label{sec:framework-hidden-priors}

The laws governing gravitational lensing are invariant under specific
transformations of some observables, whose physical features by contrast are not
\citep{gorenstein88, saha00}. This leads to parameter degeneracies when
interpreting observations and inferring physical parameters. In gravitational
lensing, all observables except for the time delay are dimensionless, and the
inference of $\kappa$ is invariant under the renormalization by an arbitrary
constant $\mu$ of the angles $\vc\theta$ and $\vc\beta$. This implies that the
inference of the $\dr{s}$'s does not depend on the radial position of the
images, but only on their relative geometrical distribution.

Another relevant degeneracy is the so called mass-sheet (or steepness)
degeneracy \citep{schneider13}, where the corresponding transformation with an
arbitrary constant $\mu$ is
\begin{equation}
	\vc\beta \longrightarrow \mu\vc\beta;\quad (1-\kappa) \longrightarrow
	\mu(1-\kappa).
	\label{eq:msd-trafo}
\end{equation}
Although the image structure remains the same, the inferred mass profile
changes, since this transformation rescales $\kappa$ by $\mu$ and adds or
subtracts the constant mass sheet $(1-\mu)$ to the lens. This implies that the
steepness of the mass profile is degenerate with the unknown source position
when exploiting a single redshift image system.

There are many ways in which one can break the mass-sheet degeneracy
\citep{saha00}. In our case, as there is little chance to measure the time
delays of the images or the source absolute magnitude, we can break the
degeneracy by again exploiting a second image system. Sources at different
redshifts imply different lens equations \eqref{eq:lens-eq}. These equations are
then no more simultaneously invariant under the transformation in
\eqnref{eq:msd-trafo}, implying that the profile steepness is constrained by
$\Xi(\z{1},\z{2},\Omega)$ in \eqnref{eq:ratios-ratio}.

Under the variation of $\Omega$, the values for the distance ratios $\dr{s}$'s
change, and consequently also the inferred steepness. In other words, when the
cosmological parameters are set free, the mass-sheet degeneracy cannot be
completely broken by exploiting multiple source redshifts. A degeneracy between
the profile steepness and the cosmological parameters $\Omega$ still remains.

The typical approach when modeling gravitational lenses is to assume parametric
models, which require a functional form for the mass profile that is in general
not invariant under the transformation in \eqnref{eq:msd-trafo}. This assumption
breaks the mass-sheet degeneracy even if the cosmological parameters are set
free and therefore leads to unwanted priors when inferring the cosmological
parameters, which may have a significant influence on the estimate of the
cosmological parameters or at least on their uncertainties.

The inner density slope may vary from cluster to cluster and the best
theoretical prediction for it is still debated \citep{limousin12,newman13}.
Thus in order to make use of parametric models in the context of cosmological
parameter determination a proper analysis should be devoted to the mass-sheet
degeneracy. Considerable effort has been made to break this degeneracy when
modelling and reconstructing the mass distribution of galaxies and galaxy
clusters from gravitational lensing observations
\citep[e.g.,][]{suyu12,collett13,greene13,umetsu13}. Making use of a
non-parametric cluster mass description \citet{bradac04,bradac05} combined weak
and strong lensing to reconstruct the cluster mass profile and break the
mass-sheet degeneracy. In the analysis of time-delay galaxies,
\citet{suyu10,suyu13} broke degeneracies by complementing lensing data with
additional information that constrains the lens mass profile, such as the
measurements of the stellar velocity dispersion or the mass distribution along
the line of sight.
	
To avoid a biased inference a detailed analysis of the lens, exploiting
observations other than strong gravitational lensing, is required in parametric
models. We take an alternative approach and consider the more flexible free-form
modeling of gravitational lenses to determine the cosmological parameters
exploiting strong lensing observations alone.

\section{Free-form lens modeling}
\label{sec:free-form}

For a general mass distribution the inversion problem of the lens equation
\eqref{eq:lens-eq} cannot be solved analytically but needs to be treated
numerically. Moreover, in free-form reconstruction of gravitational lenses one
has to deal with a much higher number of parameters than in parametric models.
Crucially, however, the relationship between $\vc\alpha$ and $\kappa$ in
\eqnref{eq:def-angle} is linear due to the weak field limit of gravitational
lensing \citep{schneider92-SEF}. We can therefore discretize the mass distribution
into grid cells, or pixels, with constant convergence $\kappa_i$ and rewrite
\eqnref{eq:lens-eq} for the $j^\mathrm{th}$ image at redshift $z_j$ as
\begin{equation}
	\Delta_j\tilde{\vc\beta}_j
	= \vc\theta_j - \Delta_j \sum_i \kinf_i \tilde{\vc\alpha}_i(\vc\theta_j) ,
	\label{eq:lens-eq-lin}
\end{equation}
where $\Delta_j\kinf_i\tilde{\vc\alpha}_i(\vc\theta_j)$ is the contribution to
$\vc\alpha$ of the $i^\mathrm{th}$ pixel, $\tilde{\vc\beta}_j = \vc\beta_j
(\Delta_j)^{-1}$, and for all images of the same source the
$\tilde{\vc\beta}_j$'s and $\Delta_j$'s are equal. This approach has been
followed by \citet{2004AJ....127.2604S} and \citet{2008ApJ...681..814C}.
Relations in the form of \eqnref{eq:lens-eq-lin} for $p$ observed images
construct a system of $2p$ linear equations
\begin{equation}
	\mat A x = b ,
	\label{eq:lin-eq}
\end{equation}
where $\mat A \in \R^{2p\times n}$, $b\in\R^{2p}$ is the constant vector of the
$\vc\theta_j (\Delta_j)^{-1}$, and $x\in\R^n$ is the vector containing the $n$
free parameters $\kinf_i$ and $\tilde{\vc\beta}_j$. This system is
underdetermined as in general $2p\ll n$. Thus the solution space of
\eqnref{eq:lin-eq} is unbounded and we need additional constraints to obtain a
non-empty compact solution set. Moreover, some of these solutions are unphysical
and have to be excluded.

Together with the constraints on the image positions, we require that the mass
profile is well-behaved, i.e.~non negative and smooth \citep{coles08}. These
simple priors limit the solution space to a finite and physical set. To still
maintain the linearity, we impose physically motivated constraints on the
problem in the form of a system of $m$ linear inequalities
\begin{equation} \mat C x \leq d , \label{eq:lin-ineq} \end{equation}
where $\mat C \in \R^{m\times n}$, and $d\in\R^m$ is a constant vector.

We take linear constraints, which impose that (i) the mass must be positive
everywhere, (ii) its variations must be smooth, and (iii) the local density
gradient must point within $45^\circ$ of the center. In addition, (iv) the
arrival time order as well as the parity of the images are considered. The
conditions (i) and (iv) are trivial requirements to ensure a positive mass
density, where the produced images are located at the correct stationary points
of the arrival time surface. The astrophysically motivated conditions (ii) and
(iii) are required to exclude solutions, which mathematically satisfy the
equations but are manifestly unphysical. The smoothness of the profile is
achieved by imposing that the density of a pixel be no more than twice the
average density of its neighbors \citep{coles08}, whereas the condition on the
local density gradient guarantees an overall decaying mass density profile.

These criteria are weak and cannot drive the derivation of the mass profile,
which is determined by the constraints on the image positions. They only ensure
the solutions space to be bounded by requiring the mass density to satisfy some
basic physical requirements. Moreover, they have been tested against either
synthetic lenses from $N$-body and hydro simulations \citep{saha06,saha09} or
toy models following NFW, power-law or isothermal profiles
\citep{sereno12,my:lubini12}. It is consistently found that as long as the
number of multiple images is large enough, the mass profile is determined by the
data alone whereas the priors have only a role in the sampling strategy. An
exhaustive discussion and a proper mathematical description of the assumed
priors can be found in \citet{coles08}.

It is important to notice that we deliberately excluded any prior constraints on
the steepness of the mass profile. Excluding such constraints is crucial to
avoid uncontrolled priors on the cosmological parameters, because the steepness
of the mass profile degenerates with cosmological parameters (see
\secref{sec:tests}).

The solution set of \eqnref{eq:lin-ineq} is then a subset of $\R^n$, which is
bounded by the $m$ hyperplanes representing the constraints. On the other hand,
the solution set of \eqnref{eq:lin-eq} is an affine space of $\R^n$ having
dimension $\dof = n - 2p$. Hence, the solution set of our problem is given by
the intersection between these two sets, which constructs a non-empty convex
polytope $S$, or simplex, embedded in the affine space. \eqnref{eq:lin-eq}
therefore serves to reduce the dimension of the problem from $n$ to $\dof$.

We are interested in finding the volume of the simplex $S$. As this is in
general not possible, we will derive the volume from an uncorrelated random
sample $X$ drawn uniformly from $S$ (see \secref{sec:method}). These parameter
spaces, however, are typically embedded in 100 or more dimensions, and therefore
the sampling becomes numerically challenging. To obtain an uncorrelated sample
$X$ of points in $S$ we use the gravitational lens modeling framework {\tt
	GLASS} \citep[Coles et al.~in prep.]{my:lubini12}, which is designed for
free-form lens modeling. {\tt GLASS} uses an MCMC method based on the
Metropolis-Hasting algorithm \citep{1953JChPh..21.1087M,1970...Hastings} with a
symmetric proposal density function, which uses \eqnref{eq:lin-ineq} to hint at
the shape of $S$. This allows efficient sampling of the parameter space $S$
despite the high-dimensionality. The algorithm gives an uncorrelated random
sample $X$, whose distribution is as good as an uniform random sample of $S$
\citep{my:lubini12}.

\section{Method}
\label{sec:method}

\subsection{Occam's razor}

The method we propose to infer the probability distribution of the cosmological
parameters in strong gravitational lensing observations employs a Bayesian
approach for model comparison. A standard method to estimate the cosmological
parameters $\Omega$ is to append them to the vector of the free parameters $x$
and fit all the parameters together by maximizing the likelihood function. In
order to be able to find a best fit, however, more data points than parameters
are needed. This is possible in the case of parametric models, as in general the
number of image position coordinates is larger than the number of free model
parameters. The problem is therefore overdetermined and cannot be solved
exactly, but the maximum of the likelihood function can be found. At a first
level of inference, assuming flat priors, the probability distribution of the
parameters $\Omega$ is then given by the likelihood function marginalized over
the other model parameters $x$.

In our case the number of parameters is much larger than the data points.
Hence, the number of solutions is infinite, i.e.~all $x \in S$ exactly solve
\eqnref{eq:lens-eq} and have the same likelihood with $\chi^2=0$. We are
therefore not able to determine the best fit parameters using Bayesian first
level inference. At the second level of inference, we can estimate the
plausibility of different models given the data, even in an underdetermined
case. This is possible because of the Bayesian Occam's razor for model
comparison \citep{mac03}, where the plausibility of a model is proportional to
the volume occupied by $S$ in the parameter space.

The Occam factor for the cosmological parameters in free-form lens modeling is
derived as follows. Let us consider different gravitational lens mass
reconstruction models $\mathcal{M}_l$ which reproduce the data $\mathcal{D}$
given by the image positions, each time assuming different values $\Omega_l$ for
the cosmological parameters. The model $\mathcal{M}_l$ is then described by the
cosmological parameters and the setting parameters defining the discretized
convergence map. The free parameters $x$ of these models are the mass of the
pixels and the source position coordinates, whereas the posterior probability of
our problem is given by Bayes' theorem as
\begin{equation}
	P(x|\mathcal{D},\mathcal{M}_l) =
	\frac{P(\mathcal{D}|x,\mathcal{M}_l)P(x|\mathcal{M}_l)}{P(\mathcal{D}|\mathcal{M}_l)}.
\label{eq:bayes-theorem}
\end{equation}
Assuming flat priors for the models, which means that $P(\mathcal{M}_l)$ is
constant, the probability of a model given the data is proportional to the
evidence $P(\mathcal{D}|\mathcal{M}_l)$ in \eqnref{eq:bayes-theorem}.
Marginalizing over $x$ we obtain
\begin{equation}
	P(\mathcal{M}_l|\mathcal{D}) \propto P(\mathcal{D}|\mathcal{M}_l) =
	\int_{\R^n} \!\! P(\mathcal{D}|x,\mathcal{M}_l) P(x|\mathcal{M}_l) \dd x.
\label{eq:bayes-evidence}
\end{equation}

On the one hand, the prior $P(x|\mathcal{M}_l)$ can be obtained considering
\eqnsref{eq:lin-eq} and \eqref{eq:lin-ineq}, and assuming that the data
$\mathcal{D}$, i.e.~the positions of the images, are unknown (see
\secref{sec:method-prob-comp}). These equations define the region
$R_l\subset\R^n$, in which $x$ is allowed \emph{a priori} by the model
$\mathcal{M}_l$ before the data arrive. Since only for $x\in R_l$ these
equations are exactly satisfied, the prior $P(x|\mathcal{M}_l)$ is uniform in
$R_l$ and vanishes outside. That means for $x\in R_l$
\begin{equation}
	P(x|\mathcal{M}_l)=1/V(R_l) ,
	\label{eq:bayes-prior}
\end{equation}
where $V(R_l)$ is the volume of $R_l$. On the other hand, since only for $x \in
S_l \subset R_l$ the data $\mathcal{D}$ are exactly reproduced by the model
$\mathcal{M}_l$, the likelihood reads
\begin{equation}
	P(\mathcal{D}|x,\mathcal{M}_l) = \left\{\begin{array}{l} 1 \quad \text{if}\ x \in S_l \\
	0 \quad \text{if}\ x \not\in S_l \end{array}\right. .
\label{eq:bayes-posterior}
\end{equation}
From \eqnsref{eq:bayes-prior} and \eqref{eq:bayes-posterior} the evidence in
\eqnref{eq:bayes-evidence} can be reduced to 
\begin{equation}
	P(\mathcal{D}|\mathcal{M}_l) = \int_{S_l} P(x|\mathcal{M}_l) \dd x =
	\frac{V(S_l)}{V(R_l)} .
\label{eq:occam-factor}
\end{equation}
This ratio is called the Occam factor and it is the ratio between the posterior
and the prior accessible volume in the parameter space \citep{mac03}. As the
purpose of this paper is to obtain confidence levels for the cosmological
parameters, we are only interested in the Bayes factor
\begin{equation}
	\frac{P(\mathcal{M}_1|\mathcal{D})}{P(\mathcal{M}_2|\mathcal{D})} =
	\frac{V(S_1) \, V(R_2)}{V(S_2) \, V(R_1)} ,
	\label{eq:bayes-factor}
\end{equation}
where the probabilities of the models assuming either $\Omega_1$ or $\Omega_2$ are
compared. Assumptions causing a small collapse of the space volume after the data
arrive, i.e.~high Occam factor, are favored compared to the one having a larger
collapse \citep{mac03}. Thus we estimate the plausibility of the parameters
$\Omega$ by means of \eqnref{eq:bayes-factor}, where the volumes of $S$ and
$R$ have to be computed.

\subsection{Probability computation}
\label{sec:method-prob-comp}

Computing the volumes of such simplices, however, is not without its own
problems, since it has been shown that computing the exact volume of a convex
polytope is \#P-hard, even if all its vertices are known \citep{dyer88}. We are
therefore not able to compute $V(S)$ in high-dimensions, because the number of
vertices has a huge combinatorial upper bound \citep{nla.cat-vn2686816}. An
approximation of $V(S)$ is therefore needed. As we are only interested in ratios
between volumes in \eqnref{eq:bayes-factor}, the volume of a simplex can be well
approximated by means of its covariance matrix $\mat\Sigma$ through
\begin{equation}
V(S) \simeq \sqrt{\det \mat{\Sigma}} = \prod_{i=1}^\dof \sqrt{\lambda_i} ,
\label{eq:volume}
\end{equation}
where $\lambda_i$ are the eigenvalues of $\mat\Sigma$. This means that we
approximate the true volume $V(S)$ with the volume of an $\dof$-dimensional
ellipsoid or hyperrectangle, whose axis lengths correspond to the square root of
the eigenvalues of $\mat \Sigma$.

To estimate $\mat\Sigma$ we use the covariance matrix $\widehat{\mat\Sigma}$ of
a sample of points $X$ uniformly and randomly distributed in $S$, which, as
detailed in \secref{sec:free-form}, we achieve using the program {\tt GLASS}. Since
the simplex $S$, and therefore also $X$, are embedded in a $\dof$-dimensional
affine space, the matrix $\widehat{\mat\Sigma}\in\R^{n \times n}$ is singular,
and all the $n-\dof=2p$ eigenvalues of $\widehat{\mat\Sigma}$, whose eigenvectors
are perpendicular to the affine space, vanish. The sample size $|X|$ has
accordingly to be $\geq \dof + 1$, and the points of $X$ must not lie on
the same $(\dof-1)$-dimensional hyperplane of $\R^n$, meaning that $\dof$
eigenvalues of $\widehat{\mat\Sigma}$ have to be strictly positive. Moreover,
to reasonably estimate $\mat\Sigma$, samples with $|X|\gg\dof$ are needed. When
too few points are used, especially in high dimensions, the product in
\eqnref{eq:volume} will have a huge statistical uncertainty due to the
randomness of the sampling.

\begin{figure}
\centering
\def\svgwidth{\columnwidth}{\footnotesize\input{p2.tex}}

\caption{The prior accessible volume $V(R)$ is shown schematically for two
	different cosmological parameter assumptions $\Om{1}$ and $\Om{2}$. Here a
	2-dimensional graph is shown for simplicity, where $x=(\kappa_i,\kappa_j)$.
	Equation $\mat C x \leq 0$ constrains the solid angle of $R$, whereas
	equation $\mat A x = b$ defines the affine solution space. The only $\Omega$
	dependent part in these two equations is the vector $b$. The volumes
	$V(R_1)$ and $V(R_2)$ are then proportional to $d_1=\Vert \mat A^{-1} \, b_1
	\Vert$ and $d_2=\Vert \mat A^{-1} \, b_2 \Vert$, since in this case
	$\dof=1$.}

\label{fig:prior}
\end{figure}
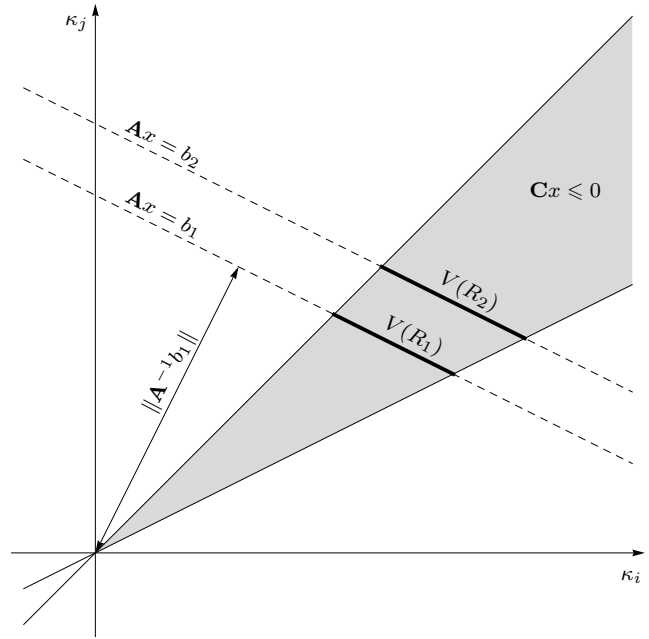

As already stated above, the prior accessible volume $V(R)$ is given considering
\eqnsref{eq:lin-eq} and \eqref{eq:lin-ineq}, and assuming $\mathcal{D}$ to
be unknown. $V(R)$ simply reflects the degeneracy in \eqnref{eq:kinf}, i.e.~the
fact that the smaller the distance ratios $\dr{s}$, the larger the value of the
convergence $\kinf$, and, finally, the larger $V(R)$. On the one hand, in
\eqnref{eq:lin-ineq} the constraints on the arrival time and the parity are
unknown, whereas the remaining constraints, which consist of the smoothness
constraints of the mass distribution, the constraints on the local gradient, and
$\kappa_i\geq0$, can all be written in the form $c \cdot x \leq 0$, where $c$ is
a constant vector. Therefore, the solution set is bounded by hyperplanes passing
through the origin of the parameter space. Hence, these constraints do not
depend on the norm $\Vert x \Vert$, which implies that the model dependent part
of \eqnref{eq:lin-ineq} constrains only the solid angle of $R$. This is shown
schematically as the gray region in \figref{fig:prior}. It is important to
notice that only the information obtained from lensing observations and not the
additional constraints of \eqnref{eq:lin-ineq} constrain the norm $\Vert x
\Vert$ and thus the mass of the pixels.

On the other hand, as the unknown data imply unknown $\vc\theta$ and $\mat A$,
the solution set of \eqnref{eq:lin-eq} is an affine space with unknown angular
position. The solid angle as well as the angular position of $R$ do not depend
on $\Omega$ and are therefore equal for all $\mathcal{M}$. The only cosmological
parameter dependent part in these equations is given by the vector $b$ in
\eqnref{eq:lin-eq}, since its components are proportional to
$\Delta_j(\Omega)^{-1}$. The distance $d$ between the affine space and the
origin of the parameter space is defined by $b$ through
\begin{equation}
	d = \Vert \mat A^{-1} \, b \Vert ,
\end{equation}
where $\mat A^{-1}$ denotes the Moore-Penrose pseudoinverse of $\mat A$.
Since $\vc\theta$ and $\mat A$ are unknown we assume random values for their
components, implying that
\begin{equation}
	d\,\mathrel{\raisebox{4pt}{$\sim$}\hspace{-7.5pt}\hbox{$\propto$}}
	\langle \Delta_j(\Omega)^{-1} \rangle,
\end{equation}
where $\langle \,\cdot\, \rangle$ is the arithmetic mean over the images $j$.
For the probability ratio in \eqnref{eq:bayes-factor} it is enough to consider the
cosmological parameter dependent part, which for the volume of $R$ is given by
\begin{equation}
	V(R) \propto d^\dof \propto \mathcal{H}(\Delta_j(\Omega_l))^{-\dof},
	\label{eq:volume-of-R}
\end{equation}
where $\mathcal{H}(\,\cdot\,)$ is the harmonic mean over the images $j$.
\figref{fig:prior} shows schematically the volumes $V(R_1)$ and $V(R_2)$
obtained considering two different affine spaces, where the vectors $b_1$ and
$b_2$ correspond to $\Omega_1$ or $\Omega_2$, respectively. The volumes are
then proportional to $d_1^\dof$ and $d_2^\dof$. In this example $\dof=1$,
since for simplicity we show a 2-dimensional plot.

Considering \eqnsref{eq:volume} and \eqref{eq:volume-of-R}, we obtain the final
estimate for the probability of a cosmological model $\mathcal{M}_l$ given the
data $\mathcal{D}$, which reduces to
\begin{equation}
	P(\mathcal{M}_l|\mathcal{D}) \propto
	\left(\prod_{i=1}^{\dof} \sqrt{\widehat{\lambda}_{l,i}}\right)\cdot
\mathcal{H}(\Delta_j(\Omega_l))^\dof,
	\label{eq:prob-omega}
\end{equation}
where $\widehat{\lambda}_{l,i}$ are the $\dof$ positive eigenvalues of
$\widehat{\mat\Sigma}_l$. To better understand this result, let us consider the special
case of a single source redshift $\z{s}$. The harmonic mean then reduces to
$\mathcal{H}(\Delta_j(\Omega_l))=\dr{s}(\Omega_l)$ and \eqnref{eq:lens-eq-lin}
can be rewritten in the variables $\kappa_i=\kinf_i\cdot\dr{s}(\Omega_l)$ and
$\vc\beta_j=\tilde{\vc\beta}_j\cdot\dr{s}(\Omega_l)$ for all the images $j$.
Thus the factor $\dr{s}(\Omega_l)$ cancels out in \eqnref{eq:lens-eq-lin}, which
does not depend anymore on the choice of $\Omega_l$. The simplex $S_l$ in the
$(\kinf_i,\tilde{\vc\beta}_j)$-space can be transformed in the corresponding
simplex $S'$ in the $(\kappa_i,\vc\beta_j)$-space simply by multiplying the
coordinates with the factor $\dr{s}(\Omega_l)$. Hence, for
\eqnref{eq:prob-omega} we obtain
\begin{equation}
	P(\mathcal{M}_l|\mathcal{D}) \propto V(S_l)\cdot\dr{s}(\Omega_l)^{\dof} = V(S') ,
\end{equation}
meaning that the probability is proportional to the volume of the simplex in the
$(\kappa_i,\vc\beta_j)$-space. By construction, $V(S')$ does not depend on the
choice of $\Omega_l$, and thus neither does $P(\mathcal{M}_l|\mathcal{D})$.
As expected from the discussion in \secref{sec:framework}, regarding the
degeneracy in \eqnref{eq:kinf}, we cannot constrain cosmological parameters
with a single source plane.

In the case of multiple source redshifts, since $\kappa_i$ depends on the
redshift of the source, there are many different $(\kappa_i,\vc\beta_j)$-spaces.
Hence, instead of $V(S')$ in \eqnref{eq:prob-omega} we have to consider an
average of the volumes of the simplices in the $(\kappa_i,\vc\beta_j)$-spaces to
be proportional to the probability. For this reason the factor
$\mathcal{H}(\Delta_j(\Omega_l))^\dof$ is required in \eqnref{eq:prob-omega},
and thus it accounts for the degeneracy in \eqnref{eq:kinf}.

\section{Tests}
\label{sec:tests}

\begin{figure*}
\centering
\includegraphics[width=\textwidth]{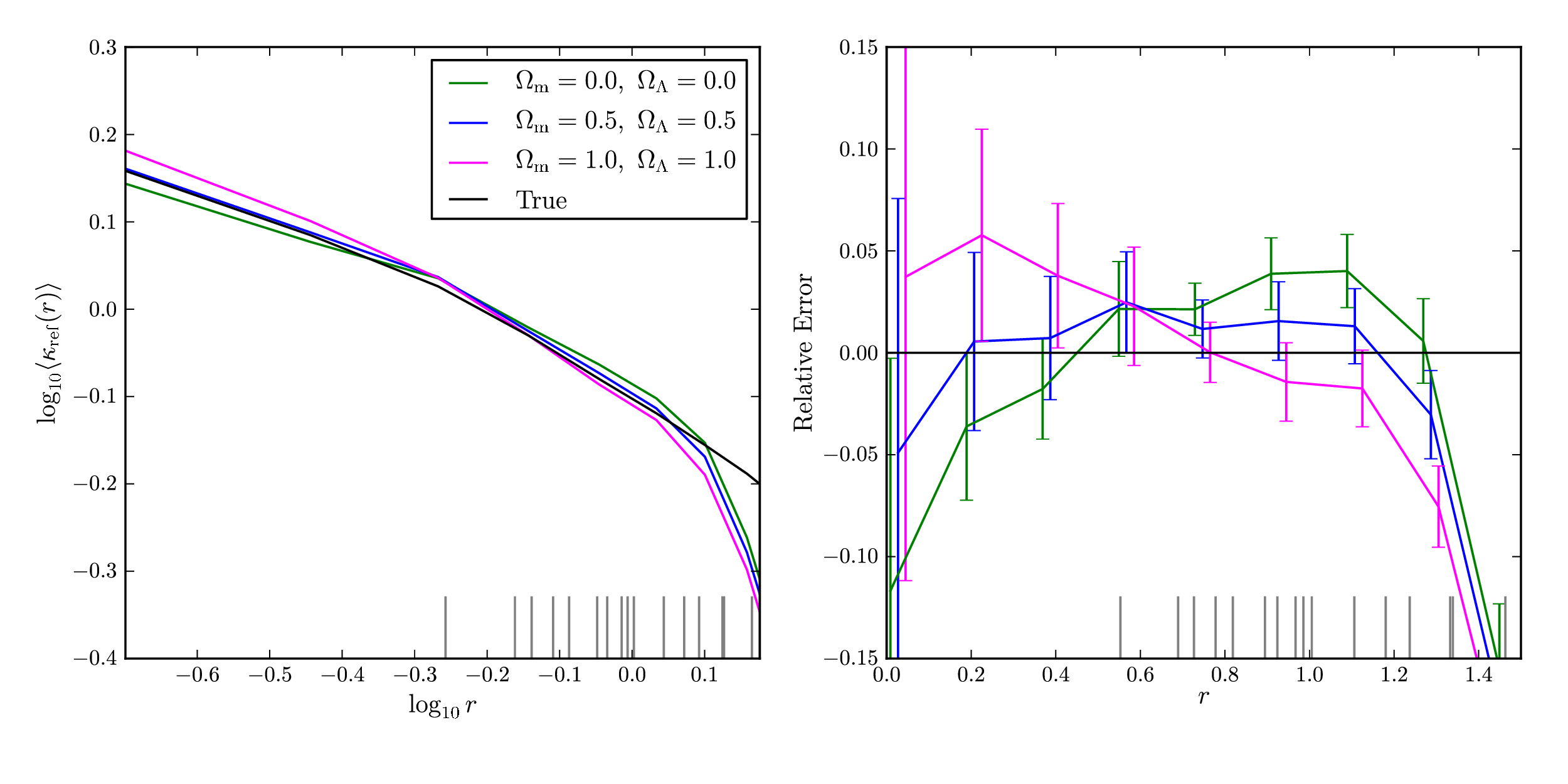}

\caption{\emph{Left panel:} The reconstructed lens profile $\kref$ ($\z{ref}=2$)
	of the same lens are shown for three different sets of cosmological
	parameters. Only the assumed values for $\Omega=(\Om{m},\Om{\Lambda})$
	change between the reconstructions. $\langle\kref(r)\rangle$ is the mean of
	$\kref$ within the pixelated ring of radius $r$, which is in units of
	$\langle\Vert\vc\theta_j\Vert\rangle$. The images, whose positions are
	marked by gray lines, are located at radii $0.55\leq r\leq1.46$, or
	$-0.26\leq\log_{10}r\leq0.16$. For comparison the true mass profile, used to
	produce the image systems under the assumption of $\Om{ref}=(0.5,0.5)$, is
	shown in black. The fits are not physical outside the outermost images,
	since there is no lensing information on the mass profile. \emph{Right
		panel:} To highlight the degeneracy between steepness and cosmological
	parameters, the relative error with respect to the true mass profile
	$\kref/\kappa_\mathrm{true}-1$ is shown. The three curves in the plot have
	been slightly shifted to avoid the overlapping of the $1\sigma$ error bars.}

\label{fig:steepness-degeneracy}
\end{figure*}

We test our method by means of synthetic lenses produced with {\tt Gravlens}
\citep{keeton01b,keeton01a}. This software builds lenses as parametric mass
distributions and finds, through numerical inversion of the lens equation
\eqref{eq:lens-eq}, the position, the arrival time, and the parity of all the
images produced by a given source position. Image configurations $\mathcal{D}$
are produced by the synthetic lenses assuming a reference set of true
cosmological parameters $\Om{ref}$. This is a realistic testing procedure, since
our method utilizes a discretized mass distribution, whereas the adopted image
configurations are obtained from smooth lenses, just as real galaxies or galaxy
clusters. For simplicity and without loss of generality we renormalize each
image position $\vc\theta$ by the mean radius $\langle\Vert\vc\theta_j\Vert
\rangle$ of all the images. This leaves the results unchanged (see
\secref{sec:framework}) and allows for an easier comparison between different
configurations and mass profiles.

For each cosmological model, by means of {\tt GLASS}, we then find a set $X$ of
discretized mass distributions and source positions, i.e.~$x\in S_l$, which
exactly reproduce the synthetic image configuration.

The discretized mass distribution is defined within a radius $R$, which is
divided into $P$ pixels \citep{2004AJ....127.2604S}. Thus the models
$\mathcal{M}_l$ depends not only on $\Omega_l$ but also on the model parameters
$R$ and $P$. While the latter changes the resolution of the discretization, $R$
does not influence $n$. We fix $P$ \emph{a priori} and marginalize over $R$,
that is
\begin{equation}
	P(\Omega_l,\mathcal{M}|\mathcal{D})=
	\int_\R P(\Omega_l,R,\mathcal{M}|\mathcal{D})\,\dd R , \label{eq:marginalization}
\end{equation}
where the dependency on $\Omega_l$ and $R$ is written explicitly. The
marginalization with respect to $R$ is important, since for different $\Omega_l$
the probabilities may have their maximum at different $R$'s. Moreover, this
procedure enables us to exclude those image configurations where
$P(\Omega_l,R,\mathcal{M}|\mathcal{D})$ as a function of $R$ is not single
peaked or heavily depends on the choice of $\Omega_l$. Without accounting for
different map radii, cosmological information would be strongly affected by the
discretization and thus no longer reliable.

To test the model, we consider a realistic scenario, where the lens located at
redshift $\z{l}=0.2$ is a massive cluster, that follows a single NFW profile
with concentration parameter $c_{200}=5$, mass $M_{200}=10^{15}M_\odot/h$, and
ellipticity $e=0.15$. To produce the synthetic image configurations in this
section we assume the $\Lambda$CDM cosmological model with
$\Om{ref}=(\Om{m},\Om{\Lambda})=(0.5,0.5)$. Results for this reference
cosmological model are then compared to two competitive models, the empty open
universe $\Om{emp}=(0,0)$ and a closed universe, $\Om{cld}=(1,1)$.

As discussed in \secref{sec:framework} the mass-sheet degeneracy is still there
even exploiting multiple source redshifts and setting $\Omega$ free. We consider
an image configuration with 5 different source planes producing overall 16
images, whose redshifts are $\z{s}=1.0,\ 1.5,\ 2.0,\ 2.5$, and $3.0$,
respectively. The mean mass profiles for the three fits are shown in
\figref{fig:steepness-degeneracy}. The profile steepness depends on the assumed
cosmological parameters. Cosmologies with a larger $\Omega_k$ need a
shallower mass profile in order to fit the given configuration. The fit with the
correct value for $\Omega$ excellently reproduces the true mass profile within
the outermost image radius. Thus {\tt GLASS} is able to reproduce the assumed
profile and produces an unbiased mass estimate \citep{my:lubini12}.

Outside the outermost image position there is no information on the profile and
the fits are unphysical. This is because in \eqnref{eq:lin-ineq} we consider
moderate constraints and let the steepness of the profile completely free to
vary in order to avoid breaking any possible degeneracy as, for instance, the
mass-sheet degeneracy. This issue, however, has no influence on our analysis, as
these pixels are not constrained by the images and thus are independent of
$\Omega_l$. Nevertheless, these pixels have to be considered, since otherwise
the approximation $V(S)\simeq(\det\widehat{\mat\Sigma})^{1/2}$ in
\eqnref{eq:volume} is no longer valid.

\begin{figure*}
\centering
\includegraphics[width=\textwidth]{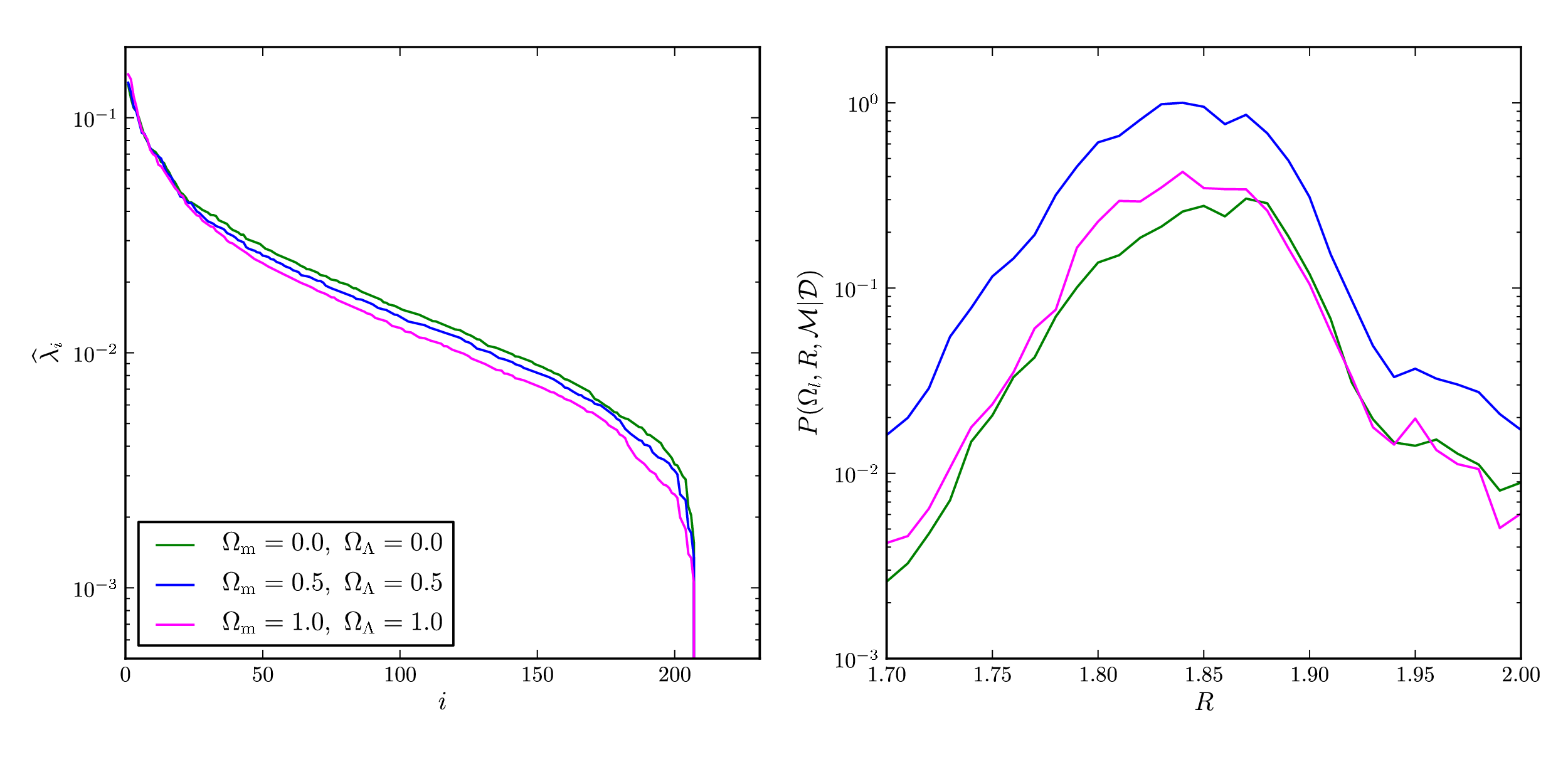}

\caption{\emph{Left panel:} The sorted list of the eigenvalues
	$\widehat{\lambda}_i$ of $\widehat{\mat\Sigma}$ for the three different fits
	with $|X|=10^4$ and $R=1.85$ are shown. The first $\dof=207$ eigenvalues are
	strictly positive, whereas the last $n-\dof=2p=24$ are not displayed, as
	they are $<10^{-14}$, which means $0$ within the machine precision.
	\emph{Right panel:} The probability $P(\Omega_l,R,\mathcal{M}|\mathcal{D})$
	as a function of $R$ is shown for the three fits. The three curves were
	rescaled in arbitrary units. The blue curve corresponding to $\Om{ref}$ is
	larger than the other two cases, meaning that this assumption is more
	likely.}

\label{fig:eigenvalues-maprad}
\end{figure*}

We elucidate and test some properties of \eqnref{eq:prob-omega} by means of an
image configuration with 3 sources producing overall $p=12$ images, i.e.~3
quads, whose source redshift are $\z{s}=1.0$, $2.0$, and $3.0$, respectively. We
produce solution sets $X$ with $|X|=10^4$ and $P=8$, which implies that there
are $n=231$ free parameters. The sorted list of the eigenvalues
$\widehat{\lambda}_i$ in the case with $R=1.85$ is shown in the left panel of
\figref{fig:eigenvalues-maprad}. Only the first $\dof = 207$ eigenvalues are
strictly positive, whereas the remaining $2p=24$ vanish, as expected. The larger
eigenvalues correspond to the more massive pixels, i.e.~those located in the
very inner region of the mass distribution. The mass-sheet degeneracy is
therefore also visible in the steepness of the curves of the three different
cosmologies.

The right panel of \figref{fig:eigenvalues-maprad} shows
$P(\Omega_l,R,\mathcal{M}|\mathcal{D})$ as a function of $R$ for the three sets
of cosmological parameters. The curves are single peaked with the maximum
around $R=1.85$ and the curve corresponding to the true cosmological parameters
has a larger normalization than the others. The marginalization over $R$ yields
the probabilities $P(\Omega_l,\mathcal{M}|\mathcal{D})$. The true cosmology is
clearly favored with $P(\Om{ref},\mathcal{M}|\mathcal{D})$ being about 3 times
larger than $P(\Om{emp},\mathcal{M}|\mathcal{D})$ and
$P(\Om{cld},\mathcal{M}|\mathcal{D})$.

We finally verified that the method is unbiased to the underlying cosmological
model of reference. Even assuming the extreme case
$\Om{ref}=(\Om{m},\Omega_\Lambda)=(2.5,0.5)$ our method could find $\Om{ref}$
within the uncertainties.

\section{Parameter determination}
\label{sec:parameter-determination}

\begin{figure*}
\centering
\includegraphics[width=\textwidth]{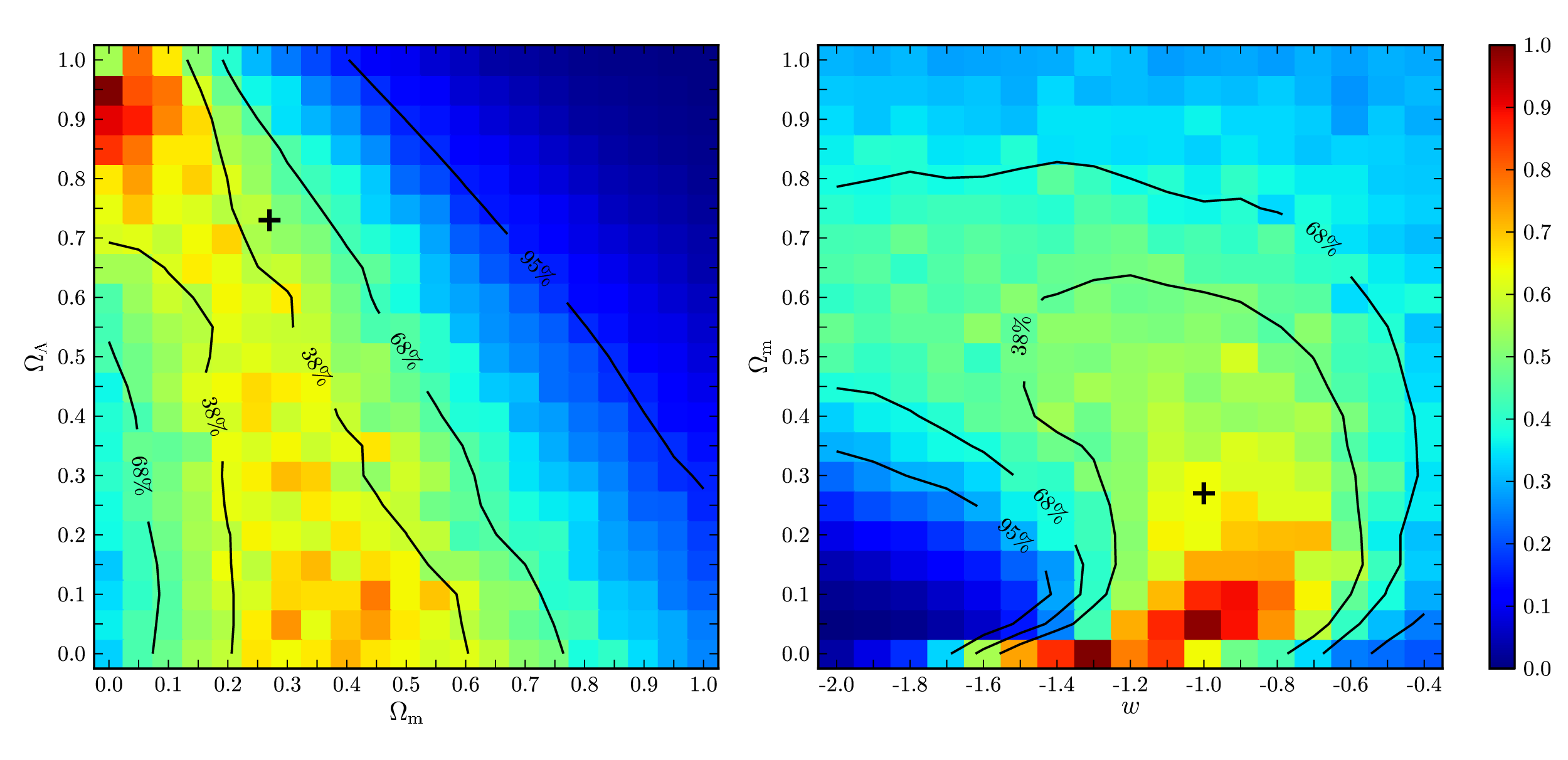}

\caption{The inferred probability distribution for a single lensing cluster in
	the $\Om{m}-\Om{\Lambda}$ plane with $w=-1$ (\emph{left panel}) and in the
	$w-\Om{m}$ plane with $\Omega_k=0$ (\emph{right panel}). The probability
	isodensity contours almost correspond to the contours shown in
	\figref{fig:om-ol-w-contours} and the area enclosed by the contours corresponds
	to respectively $38\%$, $68\%$, and $95\%$ of the total probability. The
	true values for the cosmological parameters $\Om{ref}=(0.27,0.73,-1)$ are
	shows by the black crosses.}

\label{fig:NFW-Om-Ol-w}
\end{figure*}

To show the accuracy of the method, we considered the massive cluster with
$(c_{200},M_{200},e)=(5,10^{15}M_\odot/h,0.15)$ of \secref{sec:tests} and the
concordance $\Lambda$CDM cosmology with
$\Om{ref}=(\Om{m},\Om{de},w)=(0.27,0.73,-1)$ \citep{komatsu11-WMAP7} as the
``true'' cosmological model. The synthetic image configuration is produced by 5
sources at redshifts $\z{s}=1.2$, $1.9$, $2.5$, $2.8$ and $4.0$, which produce
three quads and two doubles. The total of 16 images at different redshifts
contains the information on the cosmography.

We fit the parameters either in the $\Om{m}-\Om{\Lambda}$ plane, where
$(\Om{m},\Om{\Lambda})\in[0,1]\times[0,1]$ with $w=-1$, or in the $w-\Om{m}$
plane, where $(w,\Om{m})\in[-2,-1/3]\times[0,1]$ and $\Omega_k=0$. We divided
the two planes into grids, and computed the probabilities
$P(\Omega_l,\mathcal{M}|\mathcal{D})$ marginalizing over $R$ for each model
$\Omega_l$.

Although the sampling algorithm has been recently improved, it still has a
running time of $\mathcal{O}(n^3)$ \citep{my:lubini12}. Thus, we need to keep
the space dimension small enough to have a reasonable computation time, but
large enough such that the discretization does not compromise the fit of the
image configuration. For this reason we choose $P=8$, which corresponds to
$n=235$, and divide the planes into grids of $21\times21$ and $21\times17$
pixels, respectively.

The results are shown for the $\Om{m}-\Om{\Lambda}$ plane in the left panel of
\figref{fig:NFW-Om-Ol-w}, and for the $w-\Om{m}$ plane in the right panel. The
probability isodensity contours follow those in \figref{fig:om-ol-w-contours},
since for gravitational lensing the information of the cosmological parameters
is contained in $\Xi$. The parameter degeneracy is partially broken thanks to
the multiple source redshifts.

The method is mainly sensitive to the space curvature parameter $\Omega_k$ and
the matter density parameter $\Om{m}$ in the $\Om{m}-\Om{\Lambda}$ plane, and to
the dark energy equation of state $w$ in the $w-\Om{m}$ plane. These parameters
are almost perpendicular to the degeneracies in the respective planes, and can
therefore be inferred by means of a single lens even if with large
uncertainties.

Our method obtains an unbiased estimate of the assumed values for the
cosmological parameters within the statistical uncertainties. $\Omega_{k}$ is
retrieved with an accuracy of $0.3$ and $w$ with an uncertainty of about $0.4$,
whereas $\Om{m}$ is accurate within $0.3$.

\section{Conclusions}
\label{sec:conclusions}

Strong gravitational lensing exploiting multiple lensed background sources in
galaxy clusters is a unique tool to probe the cosmology. It relies on purely
geometrical information that does not need any calibration, and explores new
redshift ranges around $z \sim 3-4$. To extract the cosmological information
contained in lensed image systems, we exploited free-form lens modeling by means
of the framework {\tt GLASS}. This software is based on an efficient sampling
strategy that produces uncorrelated random samples \citep{my:lubini12}. These
are fundamental for our analysis, since the solution spaces we need to sample
are convex polytopes in 200 and more dimensions.

The free-form approach we investigated is more flexible than parametric
techniques, and requires only the geometrical information from strong lensing.
Parametric models demand deep knowledge of all cluster components, and assume
functional forms for the mass profiles, which break the mass-sheet degeneracy.
When inferring the cosmological parameters, however, the mass-sheet degeneracy
is still present even in the case of multiple source planes, since cosmologies
with larger $\Omega_k$ need shallower mass profiles to fit the same image
configuration. Therefore parametric models unintentionally break possible
degeneracies by adding hidden priors to the analysis. This leads to biased
estimates with unrealistically small uncertainties.

Our method does not use constraints on the steepness of the profile and accounts
for the mass-sheet degeneracy. This solves one of the main systematics in
lensing determination of cosmological parameters. The systematic effect due to
the presence of uncorrelated substructures along the line of sight
\citep{daloisio11}, which is a main source of uncertainty in cosmography, has
not been considered in this paper. However, cosmic variance plays a minor role
in the strong lensing regime.

Since in free-form modeling there are more free parameters than data points, we
cannot follow a maximum-likelihood estimation of the cosmological parameter.
Therefore, we developed a method based on Bayesian model comparison. The
probabilities are obtained through the Occam factor, which means that we take
the volume of the solution space as a tracer of the probability. We considered
the probability to be proportional to the ratio between the posterior and prior
accessible volumes.

Testing with synthetic lenses showed that our method can infer the values of the
assumed cosmological parameters. The free-form strong lensing geometrical test
we developed seems particularly promising in view of ongoing and future
observational programs. The Cluster Lensing And Supernova survey with Hubble
(CLASH) project \citep{postman12} has been deeply observing 25 massive clusters
at $0.2 \lesssim z \lesssim 0.6$. Predictions and first analyses
\citep{umetsu12, medezinski13} agree on an expected detection rate of at least
between $12$ and $15$ multiple systems per clusters. On the basis of the CLASH
clusters alone, the strong lensing test we proposed might decrease the
uncertainties of the cosmological parameters with respect to ones obtained for
one single cluster in \secref{sec:parameter-determination} by almost one order
of magnitude.

Future surveys will provide additional very large cluster samples. Euclid is
expected to detect about $5000$ clusters with prominent arcs and strong lensing
features \citep{laureijs11}. The consequent improvement of the method accuracy
over such a large sample is of one or two additional orders of magnitude. The
performance should further improve using the method in combination with
orthogonal probes such as CMB or BAO.

\bibliographystyle{mn2e}

\input{lubini.bbl}
\bsp
\label{lastpage}

\end{document}

%% file: p2.tex
\begingroup%
  \makeatletter%
  \providecommand\color[2][]{%
    \errmessage{(Inkscape) Color is used for the text in Inkscape, but the package 'color.sty' is not loaded}%
    \renewcommand\color[2][]{}%
  }%
  \providecommand\transparent[1]{%
    \errmessage{(Inkscape) Transparency is used (non-zero) for the text in Inkscape, but the package 'transparent.sty' is not loaded}%
    \renewcommand\transparent[1]{}%
  }%
  \providecommand\rotatebox[2]{#2}%
  \ifx\svgwidth\undefined%
    \setlength{\unitlength}{360bp}%
    \ifx\svgscale\undefined%
      \relax%
    \else%
      \setlength{\unitlength}{\unitlength * \real{\svgscale}}%
    \fi%
  \else%
    \setlength{\unitlength}{\svgwidth}%
  \fi%
  \global\let\svgwidth\undefined%
  \global\let\svgscale\undefined%
  \makeatother%
  \begin{picture}(1,1)%
    \put(0,0){\includegraphics[width=\unitlength]{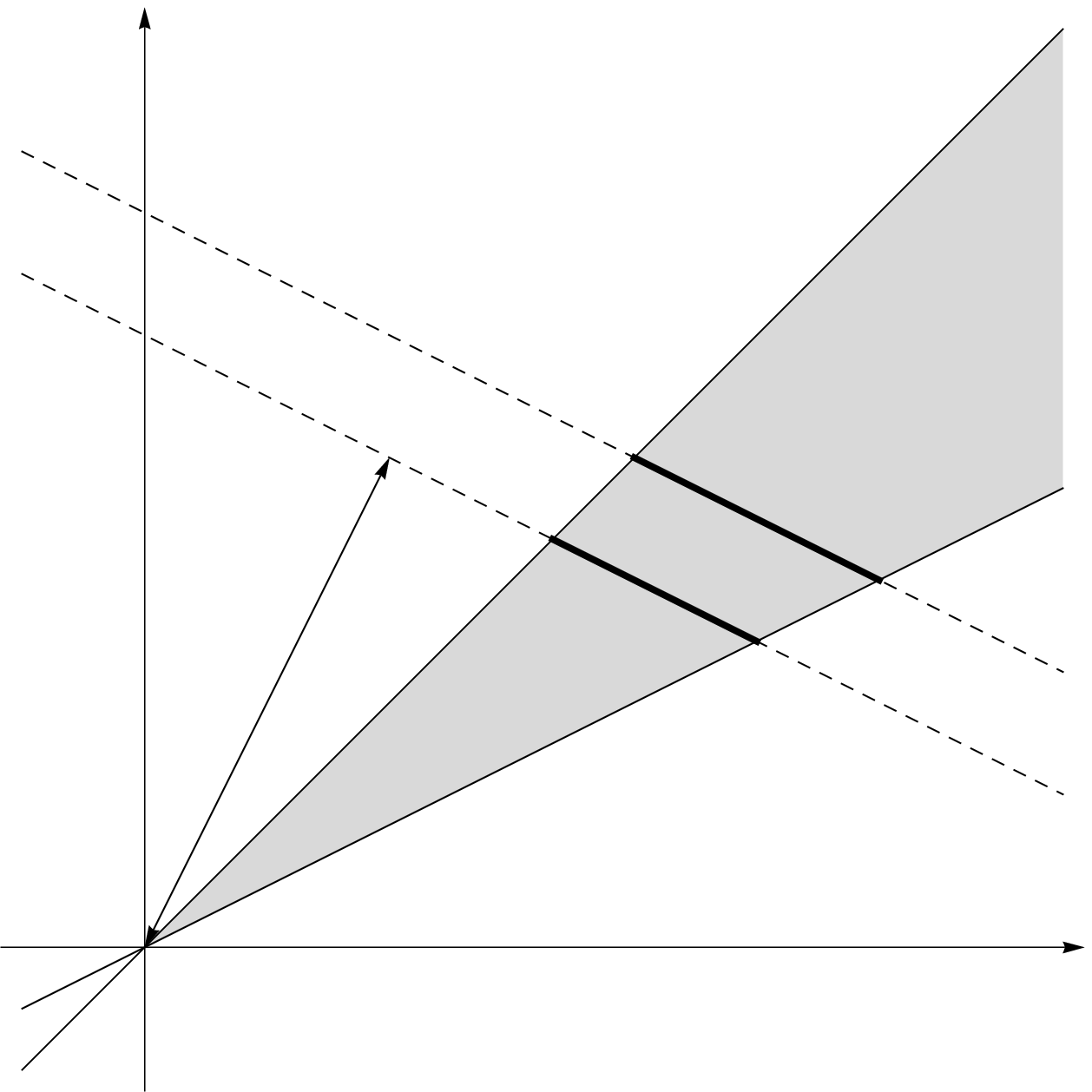}}%
    \put(0.08592591,0.96592593){\color[rgb]{0,0,0}\makebox(0,0)[lb]{\smash{$\kappa_j$}}}%
    \put(0.17882709,0.68897904){\color[rgb]{0,0,0}\rotatebox{-26.90908882}{\makebox(0,0)[lb]{\smash{$\mat Ax=b_1$}}}}%
    \put(0.17882709,0.80097904){\color[rgb]{0,0,0}\rotatebox{-26.90908882}{\makebox(0,0)[lb]{\smash{$\mat Ax=b_2$}}}}%
    \put(0.5862345,0.49046052){\color[rgb]{0,0,0}\rotatebox{-26.90908882}{\makebox(0,0)[lb]{\smash{$V(R_1)$}}}}%
    \put(0.6736419,0.56009016){\color[rgb]{0,0,0}\rotatebox{-26.90908882}{\makebox(0,0)[lb]{\smash{$V(R_2)$}}}}%
    \put(0.81777776,0.69481483){\color[rgb]{0,0,0}\makebox(0,0)[lb]{\smash{$\mat Cx\leq0$}}}%
    \put(0.22541553,0.35246692){\color[rgb]{0,0,0}\rotatebox{62.20448093}{\makebox(0,0)[lb]{\smash{$\Vert\mat A^{-1} b_1\Vert$}}}}%
    \put(0.96148139,0.09037035){\color[rgb]{0,0,0}\makebox(0,0)[lb]{\smash{$\kappa_i$}}}%
  \end{picture}%
\endgroup%